\documentclass{article}[12pt]

\begin{document}
\begin{center}
\vskip 1.0cm
{\huge \bf Self-organized critical topology \\ of stock markets}
\vskip 1.0cm
{\large \bf N.Vandewalle$^1$, F.Brisbois$^1$ and X.Tordoir$^2$}
\vskip 0.2cm

$^1$ GRASP, Institut de Physique B5, Universit\'e de Li\`ege, B-4000 Li\`ege, Belgium.

$^2$ IPNE, B15, Universit\'e de Li\`ege, B-4000 Li\`ege, Belgium.

\end{center}
\vskip 2.0cm
{\bf 
We have analyzed the cross-correlations of daily fluctuations for $N=6\,358$ US stock prices during the year 1999. From those $N(N-1)/2$ correlations coefficients, the Minimum Spanning Tree (MST) has been built. We have investigated the topology exhibited by the MST. Eventhough the average topological number is $\langle n \rangle \approx 2$, the variance $\sigma$ of the topological distribution $f(n)$ diverges! More precisely, we have found that $f(n) \sim n^{-2.2}$ holding over two decades. We have studied the topological correlations for neighbouring nodes: an extremely broad set of local configurations exists, confirming the divergence of $\sigma$. }

\newpage

In a series of works \cite{mantegna,book}, Mantegna and coworkers have studied the Minimum Spanning Tree (MST) of financial data, mainly the cross-correlations between $N$ different stocks. Building a MST is a common method in spin glasses for putting into evidence the primary structure of a complex system with a non-trivial dynamics \cite{spins}. The MST puts into evidence the cooperative behaviours in the stock market. 

Herein, we report some analysis of the topology exhibited by the MST. Due to Initial Public Offerings (IPOs) and mergers, it is difficult to perform some MST-analysis over long periods. The number of companies is indeed always evolving. We focussed our work on $N=6\,358$ US companies during the year 1999. Those stocks are all traded on the Nasdaq, NYSE and AMEX places. Cross-correlations $\rho_{ij}$ for pairs $ij$ of stocks are computed as
\begin{equation}
\rho_{ij}=\frac{<Y_iY_j>-<Y_i><Y_j>}{\sqrt{(<Y_i^2>-<Y_i>^2)(<Y_j^2>-<Y_j>^2)}}
\end{equation} where the $Y$'s are the daily fluctuations of the log of the stock prices. The MST has been built following the Kruskal's algorithm \cite{algo} in order to find out the $N-1$ most important correlated pairs of stocks among the $N(N-1)/2=20\,208\,903$ possible pairs! Figure 1 presents three parts of the MST: AMZN (Amazon.com Inc.), YHOO (Yahoo! Inc.), KO (Coca-Cola Company) and their nearest neighbours. AMZN is connected to EBAY (eBay Inc.), CMGI (CMGI Inc.) and SOFN (SoftNet Systems Inc.). YHOO is connected to ARBA (Ariba Inc.) and also connected to CMGI such that YHOO is the next nearest neighbour of AMZN. KO is connected to PEP (PepsiCo, Inc.), CCE (Coca-Cola Enterprises, Inc.) and AVP (Avon Products Inc.). In addition to KO, AVP is connected to 42 others stocks! 

One can observe a clustering of companies being in the same sector such as energy, technology, internet, transportation, food and cosmetics, etc. The clustering in various sectors, each being a branch of the MST, has been previously observed in \cite{mantegna}. In addition to the sectorization, we have observed a trend for a global clustering of stocks traded on different market places. Most Nasdaq stocks are located around the QQQ ticker (Nasdaq Composite ticker), and largest stocks of NYSE are located around DIA (Dow Jones ticker). 

One could distinguish three types of topological configurations for companies: (i) important nodes, (ii) links, and (iii) dangling ends. The important nodes (like AVP) should correspond to companies controlling or mediating the fluctuations of their neighbourhood. The links (like YHOO) are mediating the information (fluctuation) along the branches. The dangling ends (like PEP) are supposed to be the less influent companies, or to be the less influenced by the market fluctuations (like indices).

From the mathematical point of view, we can associate a ``topological number" $1 \le n \le (N-1)$ at each node of the tree. The topological number $n$ is defined as the number of connections starting from the considered site, i.e. the coordination number of that site. The statistics of the tree is thus determined by the distribution of topological numbers $f(n)$. The mean topological number is given by
\begin{equation}
\langle n \rangle = \sum_{n=1}^{\infty} n f(n).
\end{equation} If one multiplies both sides of this equation by $N$, the summation becomes equivalent to count twice all the $N-1$ connections along the tree. As a consequence, one has
\begin{equation}
\langle n \rangle = 2 {N-1 \over N} 
\end{equation} which is approximately equal to 2 when the number of sites $N$ becomes large, i.e. in our case. The above mathematical property is quite general and holds for all types of trees. In the following, we will consider that $\langle n \rangle = 2$. Thus, the first moment $\langle n \rangle$ of $f$ cannot give any information about the topology of trees since it does not reflect the shape of the tree. However, the second moment $\sigma$ can be quite useful. Let us assume a topological correlation in between neighbouring nodes. The Aboav law, a well known law in the Physics of foams \cite{aboav}, stipulates that topological correlations can be captured by the following relationship 
\begin{equation}
m_n = A + {B \over n}
\end{equation} where $m_n$ is the mean topological number of the nearest neighbours of a node characterized by a topological number $n$. $A$ and $B$ are positive constants to be determined. The form of the Aboav law implies that important nodes with large $n$ are more favorably connected to dangling ends. Indeed, when the coordination number becomes large and reaches $n \rightarrow N-1$, which correspond to a star structure, one should obtain $m_n=1$. This leads to the constraint \footnote{This constraint has a counterpart $A=5$ in the case of foams \cite{aboav}.} $A=1$. For a random structure, one expects that the statistical average of the product $n m_n$ is similar to $\langle n^2 \rangle$, one has 
\begin{equation}
\langle n \rangle + B = \langle n m_n \rangle \approx \langle n^2 \rangle = \sigma^2 + 4
\end{equation} and thus $B = \sigma^2 + 2$ and the Aboav law becomes
\begin{equation}
m_n = 1 + {\sigma^2 +2 \over n} 
\end{equation} for any kind of tree with a large number $N$ of nodes. The above equation is valid in the hypothesis of random trees. 

Here, we have defined some mathematical tools ($f(n)$, $\sigma$ and the Aboav law) for describing the topology of trees in full generality. Let us investigate the specific topology of financial MST. The log-log plot of the topological distribution is shown in Figure 2. The distribution $f(n)$ is broad and looks like a power law \begin{equation}
f(n) \sim n^{-\alpha} 
\end{equation} holding over two decades with an exponent $\alpha = 2.2 \pm 0.1$. This result means that the variance of the $f(n)$ distribution diverges even the mean $\langle n \rangle$ remains finite (see below)! In other words, nodes with high topological numbers are not so rare. For random trees, one expects an exponential decay of $f(n)$ and quite different statistics. The market is naturally self-organized in a coordination invariant structure, i.e. a self-organized critical structure \cite{bak}! One should also remark that our results have an important graphical consequence: large $n$ values for nodes are not so rare such that it is extremely hard to draw the entire MST. 

The power law (7) and the divergence of $\sigma$ are not numerical artefacts. Indeed, we have checked the Aboav-like law for which the variance of the topological distribution is relevant (see above). Figure 3 presents the  $(m_n,{1 \over n})$ diagram in which the Aboav law should be a line with a slope $\sigma^2 +2$. However, the $N$ values of the measured $m_n$ on the MST are sparsely distributed in the diagram such that no relevant linear fit can be performed. Eq.(6) is not valid expressing that statistical averages cannot be taken as for usual random topological structures. We argue that the extremely broad dispersion of local configurations in $(m_n,{1 \over n})$ expresses the divergence of $\sigma$. 

We put into evidence the emergence of some order out of the apparent disorder of the stock markets! Our results are robust. We did the same analysis for subsets of the data and we obtained the same behaviour (a power law for $f(n)$ and a divergence of $\sigma$). Also, we did the same for the ``anti-tree", i.e. the tree for the strongest anticorrelations, and we have found similar results with the same exponent $\alpha \approx 2.2$. As found in \cite{book}, the connections between pairs of stocks are mostly unchanged by taking other time periods. Of course, IPOs ($N \rightarrow N+1$) and mergers ($N \rightarrow N-1$) would influence a little the MST but we stress that the topological properties of the MST would not dramatically change.

\vskip 1.0cm
{\noindent \Large Acknowledgements}

NV thanks the FNRS for financial support. Valuable discussion with R.N.Mantegna is gratefully acknowledged.

\vskip 1.0cm
Any correspondance should be sent to Dr. Nicolas Vandewalle.

\vskip 1.0cm
{\noindent \Large Figure Captions}
\vskip 0.5cm

Figure 1 --- Three typical configurations of the nearest neighborhood of three companies on the MST: (a) AMZN (Amazon.com Inc.) connected to EBAY (eBay Inc.), CMGI (CMGI Inc.) and SOFN (SoftNet Systems Inc.); (b) YHOO (Yahoo! Inc.) is connected to ARBA (Ariba Inc.) and CMGI; (c) KO (Coca-Cola Company) is connected to PEP (PepsiCo, Inc.), CCE (Coca-Cola Enterprises, Inc.) and AVP (Avon Products Inc.). The topological indices $n$ are given. 
\vskip 0.5cm

Figure 2 --- Log-log plot of the topological distribution $f(n)$ of the MST. The continuous line is a fit with the power law (7). 
\vskip 0.5cm

Figure 3 --- The $(m_n,{1 \over n})$ diagram for the $N=6\,358$ companies. No Aboav (linear) behaviour emerges. 
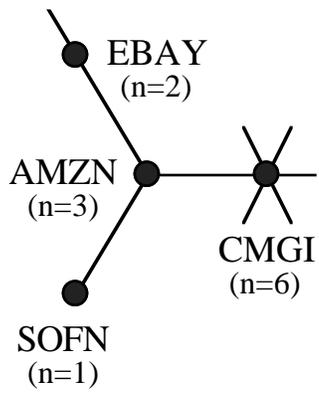
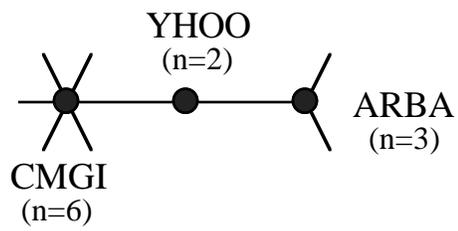
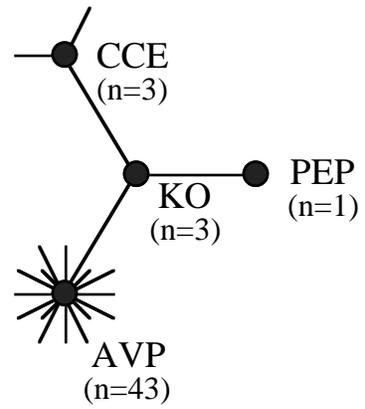
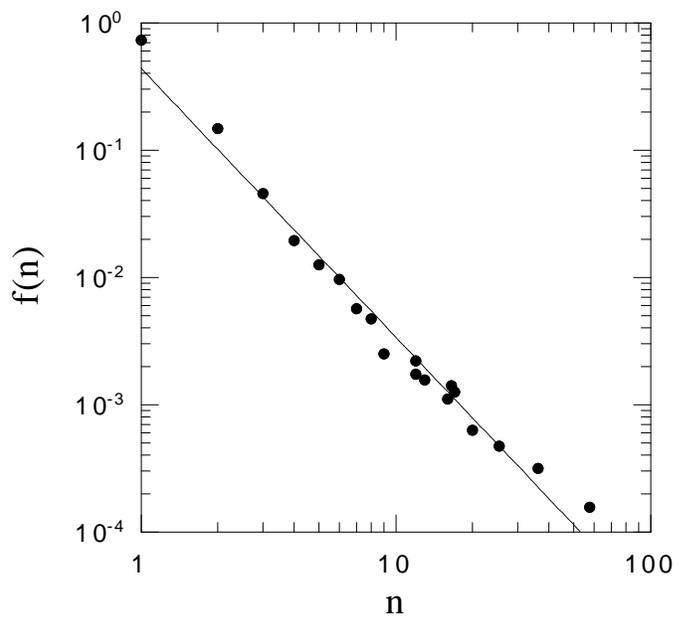
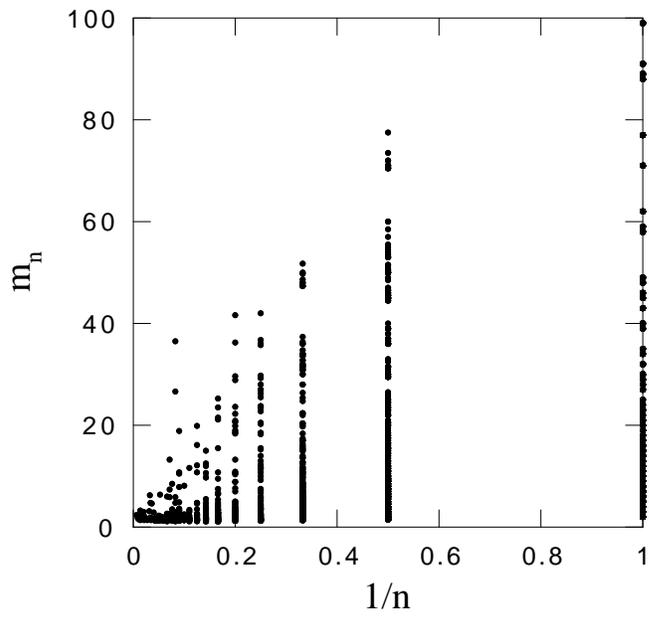
\end{document}